\documentstyle[amssymb,aps]{revtex}
\begin{document}
\title{An one-time-pad key communication protocol with entanglement}
\author{Qing-yu Cai}
\address{Wuhan Institute of Physics and Mathematics, The Chinese Academy of\\
Sciences,Wuhan, 430071, People's Republic of China \\
and Graduate School of the Chinese Academy of Sciences}
\maketitle

\begin{abstract}
We present an one-time-pad key communication protocol that allows secure
direct communication with entanglement. Alice can send message to Bob in a
deterministic manner by using local measurements and public communication.
The theoretical efficiency of this protocol is double compared with BB84
protocol. We show this protocol is unconditional secure under arbitrary
quantum attack. And we discuss that this protocol can be perfectly
implemented with current technologies.
\end{abstract}


It has been showed that quantum mechanics can be used to implement quantum
key distribution (QKD) [1,2]. However, this QKD idea does not exploit the
full potential of quantum mechanics. Instead of creating a secret key,
quantum mechanics can be used to send a secret message directly from Alice
to Bob [3,4,5]. But the implement of the protocol in Ref. [3] required the
use of a more than two dimensional Hilbert space on the single photon. In
Ref.[4], a ping-pong protocol has been presented which allows secure QKD and
quasisecure direct communication. However, this ping-pong protocol is not
secure under the measurement attack [6]. In this paper, we introduce an
one-time-pad key protocol that is an successful implementation for secure
direct communication using only local measurements and classical
communication with Einstein-Podosky-Rosen(EPR) pair.

As is well known that two qubits can be entangled in one of the Bell states 
\begin{equation}
|\phi ^{+}>_{AB}=\frac{1}{\sqrt{2}}(|0>_{A}|0>_{B}+|1>_{A}|1>_{B})=\frac{1}{%
\sqrt{2}}(|+>_{A}|+>_{B}+|->_{A}|->_{B}),
\end{equation}
\begin{equation}
|\phi ^{-}>_{AB}=\frac{1}{\sqrt{2}}(|0>_{A}|0>_{B}-|1>_{A}|1>_{B})=\frac{1}{%
\sqrt{2}}(|+>_{A}|->_{B}+|->_{A}|+>_{B}),
\end{equation}
\begin{equation}
|\psi ^{+}>_{AB}=\frac{1}{\sqrt{2}}(|0>_{A}|1>_{B}+|1>_{A}|0>_{B})=\frac{1}{%
\sqrt{2}}(|+>_{A}|+>_{B}-|->_{A}|->_{B}),
\end{equation}
\begin{equation}
|\psi ^{-}>_{AB}=\frac{1}{\sqrt{2}}(|0>_{A}|1>_{B}-|1>_{A}|0>_{B})=\frac{1}{%
\sqrt{2}}(|->_{A}|+>_{B}-|+>_{A}|->_{B}),
\end{equation}
where $|0>$ and $|1>$ are up and down eigenstate of the $\sigma _{z}$, and $%
|+>=\frac{1}{\sqrt{2}}(|0>+|1>),$ $|->=\frac{1}{\sqrt{2}}(|0>-|1>)$. It is
well known that such states can be used to establish nonlocal correlations
over a spacelike interval, but these correlations cannot be used for
superluminal communication [7]. Suppose Bob has an EPR pair in the single
state $|\psi ^{-}>$. He sends one qubit, say A, to Alice and keeps another.
Alice performs a local measurement on the qubit A in the basis $%
B_{Z}=\{|0>,|1>\}$. Thus this single state will collapse immediately to a
product state and the entanglement does not exist any longer. Bob also
performs a local measurement in the basis $B_{z}$. If Alice's measurement
outcome is $|0>$ ($|1>$), then she knows that Bob's measurement outcome is
certain to $|1>$ ($|0>$). But what outcome Alice obtains is random with a
probability $p=0.5$ for $\rho _{A}:=tr_{B}\{|\psi ^{-}><\psi ^{-}|\}=\frac{1%
}{2}|0><0|+\frac{1}{2}|1><1|$. To successfully transmit message from Alice
to Bob, a reliable public channel is required. After Alice received the
travel qubit, she performs a measurement on this qubit in the basis $B_{z}$.
Afterwards she sends a signal (one classical bit) based on her measurement
outcome to Bob through public channel. Bob also performs a measurement in
basis $B_{z}$ after he received Alice's signal. Assume Alice's measurement
outcome is $|0>$, i.e. she knows that Bob's succedent measurement outcome is 
$|1>$. When she want to sends a logical `0' to Bob, she `say no' to Bob
through public channel. Else, Alice `say yes' to Bob to encode a logical
`1'. If Alice's measurement is $|1>$, she can also encode the logical `0'
and `1' using the method described above. Therefore, Alice can sends her
secret message to Bob in the way described above. By default, Alice and Bob
are in message mode and communicate the way described above. With
probability $c$, Alice switches to control mode. $Control$ $mode$. After
received the travel qubit, Alice performs a measurement randomly in the
basis $B_{z}$ or $B_{x}=\{|+>,|->\}$. Using public channel, she sends the
result to Bob. Then Bob also switches to control mode. He performs a
measurement in the same basis as Bob used. Bob compares both of the results.
If both results coincide, Bob knows that there is an Eavesdropper-Eve in
line and stops the communication. Else, this communication continues. This
protocol can be described explicitly like this: (1) Bob prepares two qubits
in the Bell state $|\psi ^{-}>$. (2) Bob sends one qubit to Bob and keeps
another. (3) Alice receives the travel qubit. With probability $c$, she
switches to the control mode (4c). Else, she performs message mode (4m).
(4c) Alice performs a measurement on her qubit randomly in the basis $B_{x}$
or $B_{z}$ and tells the result to Bob through public channel. Bob then
performs a measurement in the same basis Alice used. If he finds both
results coincide, then he stops the communication. Else this communication
continues. (4m) Bob performs a measurement in the basis $B_{z}$. Then she
sends her signal to Bob through public channel. Bob also performs a
measurement in the basis $B_{z}$. When Alice want to transmit a logical `0'
to Bob, if her measurement outcome is $|1>$, she `say yes' to Bob through
public channel. Else, she `say no' to Bob. When Alice want to transmit a
logical `1' to Bob, she `say yes' to Bob with her measurement outcome $|0>$.
Else she `say no' to Bob with her measurement outcome $|1>$.(5) When all of
Alice's information is transmitted, this communication is successfully
terminated.

$Security$ $proof$. From Eq.(1)-Eq.(4), we know that when Alice and Bob
share the state $|\phi ^{+}>$, both of Alice's and Bob's results will
coincide no matter which basis they use in control mode. When Alice and Bob
share the state $|\phi ^{-}>$, their measurement outcome will coincide when
they use the measurement basis $B_{z}$. Alice and Bob will obtain the same
outcome with the measurement basis $B_{x}$ when they share the state $|\psi
^{+}>$. If and only if they share the state $|\psi ^{-}>$, their measurement
outcome will never coincide. Since Alice selects the measurement basis
randomly in control mode, the detection probability is at least $d\geq 0.5$
when the state they shared is one of $\{|\phi ^{\pm }>,|\psi ^{-}>\}$. In
order to be practical and secure, a QKD scheme must be based on existing-or
nearly existing-technology, but its security must be guaranteed against an
eavesdropper whose technology is limited only by the laws of quantum
mechanics [8]. In this protocol, Eve's aim is to find out what Bob's
measurement outcome is. Eve has no access to Alice's qubit. All that she can
do is to attack Bob's qubit in the quantum channel. For Eve, the qubit Bob
sends to Alice is in a complete mixed state which is completely
indistinguishable without measurement. Eve can use all technique allowed by
quantum laws to attack the travel qubit. The most general attack operation $%
\varepsilon $ can be described as a $completely$ $positive$ $map$ [9], which
implies that it is possible to find out an ancilla, E, initially in a pure
state $|e><e|$ uncorrelated with the system, and a unitary operator $U$,
obtaining 
\begin{equation}
\rho =\varepsilon (|\psi ^{-}><\psi ^{-}|)=tr_{E}\{U|\psi ^{-}><\psi
^{-}|\otimes |e><e|U^{\dagger }\}.
\end{equation}
After Eve's attack, the fidelity [10] of the states $|\psi ^{-}>$ and $\rho $
is 
\begin{equation}
F(|\psi ^{-}>,\rho )=tr\sqrt{<\psi ^{-}|\rho |\psi ^{-}>|\psi ^{-}><\psi
^{-}|}=\sqrt{<\psi ^{-}|\rho |\psi ^{-}>}.
\end{equation}
Let us assume that 
\begin{equation}
F(|\psi ^{-}>,\rho )^{2}=1-\gamma ,
\end{equation}
where $0\leq \gamma \leq 1$. Because of the completeness of the Bell's
states, it has that the detection probability $d$ is approximately dependent
on the quantity $\gamma $, $d\geq \gamma /2$. Let us consider that
information Eve can gain when the fidelity of the state $|\psi ^{-}>$ and $%
\rho $ is $\sqrt{1-\gamma }$. The mutual information Eve can gain from $\rho 
$ is bounded by Holevo quantity, $\chi (\rho )$ [9]. From 
\begin{equation}
\chi (\rho )=S(\rho )-\sum_{i}p_{i}S(\rho _{i}),
\end{equation}
we know that $S(\rho )$ is the $upper$ $bound$ of the information Eve can
gain from $\rho $. Nevertheless, the entropy of $\rho $ is bounded above by
the entropy of a diagonal density matrix $\rho _{\max }$ with diagonal
entries $1-\gamma $, $\frac{\gamma }{3}$, $\frac{\gamma }{3}$, $\frac{\gamma 
}{3}$. The entropy of $\rho _{\max }$ is 
\begin{equation}
S(\rho _{\max })=-(1-\gamma )\log _{2}(1-\gamma )-\gamma \log _{2}\frac{%
\gamma }{3}.
\end{equation}
Then the quantity $S(\rho _{\max })$ is the upper bound of the information
Eve can gain.

Let us consider the relation between the fidelity $F(|\psi ^{-}>,\rho )$ and
the detection probability $d$. When $\gamma =0$, i.e., $S(\rho _{\max })=0$,
information Eve can gain is zero. The detection probability $d$ is also zero
in this condition. If Eve want to gain some of Alice's information, it has $%
S(\rho _{\max })>0$. When $S(\rho _{\max })$ reaches its maximal value $%
S(\rho _{\max })=2$, it has $\gamma =\frac{3}{4}$. Since the von Neumann
entropy $S(\rho _{\max })$ is a continuous function of $\gamma $, any
information Eve gain will make her face a nonzero risk to be detected. The
more information Eve wants to gain, the bigger risk Eve has face. One bit
information obtained from $\rho $ will make him face a risk to be detected,
approximately $d\geq \frac{\gamma }{2}\thickapprox 9.5\%$. The information $%
I_{0}(d)$ Eve can gain in every message run is approximately dependent on
the detection probability $d(I_{0})$.

$Practical$ $feasibility$. Experimental quantum key distribution was
demonstrated for first time by Bennett $et$ $al$. [11]. Today, several
groups have shown that quantum key distribution is possible, even outside
the laboratory [12]. In principle, any two-level quantum system could be
used to implement quantum cryptography. In practice, photons are the optimal
systems for entanglement distribution because they propagate fast and can
preserve their coherence over long distances. Recently, given impure EPR
pairs shared between two distant communicators, physical scientists can
apply local operations and classical communication to distill a smaller
number of higher fidelity EPR pairs in a procedure known as entanglement
purification [13,14].

In the proposal for long-distance quantum communication [15], entanglement
between atomic ensembles is established by the detection of a single photon
that could have been emitted by either ensemble. In Ref.[16], Simon and
Irvine proposed a scheme where distance trapped ions are entangled by the
joint detection of two photons, one coming from each ion. Actually, only by
using a beam splitter and two photon detectors, two ions can be
probabilistically prepared in the state $|\psi ^{-}>$ [17]. And the
entangled ions can be stored perfectly in each ion trap comparing with the
photon storage. The entanglement time can be prolonged with the
long-lifetime of excited state of ion [18]. Comparing with today's
technologies, our protocol can be realized not only in principle but also in
practice.

$Discussion$ $and$ $Conclusion$. Virtually, our protocol is an one-time-pad
key protocol. In fact, this protocol is a variant of the Ekert91 [2] QKD
protocol. It also have been mentioned in a two-step quantum direct
communication protocol [19] that Alice and Bob can establish a common
one-time-pad key. It even could be considered a variant of the ping-pong
protocol [4], obtained by replacing Bob's final measurement by two
measurements performed by Alice and Bob on their respective qubits, followed
by classical communication. However, our protocol allows secure direct
communication only using local measurement and classical communication. And
it does not need a Bell-basis measurement in our protocol [19,20].

The theoretical efficiency of a QKD protocol is defined as $\varepsilon
=b_{s}/(q_{t}+b_{t})$ [21], where, for every step, $b_{s}$ is the expected
number of secret bit transmitted, $q_{t}$ is the number of qubits exchanged
on the quantum channel, and $b_{t}$ is the number of the announced bits. In
BB84 protocol, 0.5 $b_{s}$ bit transmitted needs 1 $q_{t}$ and 1 $b_{t}$. In
our protocol, 1 $q_{t}$ and 1 $b_{t}$ can help the communicator to exchange
1 secret bit. Thus, our protocol has high efficiency compared with BB84
protocol. Moreover, in this protocol, the control parameter $c$ is
determined by the communicators. It can be modulated according to the
variant quantum channel.

In summary, we propose an one-time-pad key communication protocol with
entanglement, which allows secure direct communication by using local
measurements and classical communication. This direct communication between
Alice and Bob is more efficient compared to Ekert protocol and BB84
protocol. And this protocol can be perfectly implemented with current
technologies.

\section{references}

[1] C. H. Bennett and G. Brassard, 1984, in $proceedings$ $of$ $the$ $IEEE$ $%
International$ $Conference$ $on$ $Computers$, $Systems$ $and$ $%
%TCIMACRO{\func{Si}}
%BeginExpansion
\mathop{\rm Si}%
%EndExpansion
gnal$ $\Pr oces\sin g$, Bangalor, India, (IEEE, New York), pp. 175-179.

[2] A. Ekert, Phys. Rev. Lett. 67, 661 (1991).

[3] A. Beige, B.-G. Englert, C. Kurtsiefer, and H. Weinfurter, Acta Phys.
Pol. A 101, 357 (2002).

[4] K. Bostr\"{o}m and T. Felbinger, Phys. Rev. Lett. 89, 187902 (2002).

[5] Q.-y Cai and B.-w Li, Chin. Phys. Lett. 21(4),601 (2004).

[6] Q.-y. Cai, Phys. Rev. Lett. 91, 109801 (2003).

[7] C. H. Bennett, Phys. Rev. Lett. 69, 2881 (1992).

[8] G. Brassard, N. Lutkenhaus, T. Mor and B. C. Sanders, Phys. Rev. Lett.
85, 1330 (2000).

[9] M. A. Nielsen and I. L. Chuang, Quantum computation and Quantum
Information (Cambridge University Press, Cambridge, UK, 2000).

[10] C. A. Fuchs, arXiv: quant-ph/9601020; H. Barnum, C. M. Caves, C. A.
Fuchs, R. Jozsa, and B. Schumacher, Phys. Rev. Lett. 76, 2818 (1996).

[11] C. H. Bennett, F. Bessette, G. Brassard, L. Salvail, and J. Smolin, J.
Cryptology 5, 3 (1992).

[12] N. Gisin, G. Ribordy, W. Tittel, and H. Zbinden, Rev. Mod. Phys. 74,
145 (2002).

[13] C. H. Bennett, D. P. DiVincenzo, J. A. Smolin, W. K. Wooters, Phys.
Rev. A 54, 3824 (1996).

[14] J.-W. Pan, S. Gasparonl, R. Ursln, G. Welhs and A. Zellinger, Nature
(London) 423, 417 (2003).

[15] L.-M. Duan, M. D. Lukin, J. I. Cirac and P. Zoller, Nature (London)
414, 413 (2001).

[16] C. Simon and W. Irvine, Phys. Rev. Lett. 91, 110405 (2003).

[17] S. L. Braunstein and A. Mann, Phys. Rev. A 51, R1729 (1995).

[18] X. Zhang, H. Jiang, J. Rao, and B. Li, Phys. Rev. A 68, 025401
(2003);C. F. Roos $et$ $al$.,Phys. Rev. Lett. 92, 220402(2004).

[19] F.-G. Deng, G.-L. Long and X.-S. Liu, Phys. Rev. A 58, 042317 (2003).

[20] Q.-y. Cai and B-w. Li, Phys. Rev. A 69, 054301 (2004).

[21] A. Cabello, Phys. Rev. Lett. 85, 5635(2000).

\end{document}